\documentclass{article}
\usepackage{spconf,amsmath,graphicx}
\usepackage[normalem]{ulem}
\useunder{\uline}{\ul}{}
\usepackage{multirow}
\usepackage{makecell}
\usepackage{pifont}
\usepackage{diagbox}
\usepackage{subfigure}
\usepackage{algorithm}
\usepackage{listings}
\usepackage{enumitem}
\usepackage{amssymb} 
\usepackage{bm}

\usepackage [dvipsnames] {xcolor}
\usepackage{url}

\title{SSL-Net: A Synergistic Spectral and Learning-Based Network \\ for Efficient Bird Sound Classification}

\name{Yiyuan Yang \qquad Kaichen Zhou \qquad Niki Trigoni \qquad Andrew Markham}
\address{Department of Computer Science, University of Oxford, United Kingdom}

\begin{document}
\maketitle

\begin{abstract}
Efficient and accurate bird sound classification is of important for ecology, habitat protection and scientific research, as it plays a central role in monitoring the distribution and abundance of species. However, prevailing methods typically demand extensively labeled audio datasets and have highly customized frameworks, imposing substantial computational and annotation loads. In this study, we present an efficient and general framework called SSL-Net, which combines spectral and learned features to identify different bird sounds. Encouraging empirical results gleaned from a standard field-collected bird audio dataset validate the efficacy of our method in extracting features efficiently and achieving heightened performance in bird sound classification, even when working with limited sample sizes. Furthermore, we present three feature fusion strategies, aiding engineers and researchers in their selection through quantitative analysis.
\end{abstract}


\section{Introduction}
\label{sec:Introduction}

There is a significant and increasing need for automated pipelines for bird sound classification, as these techniques hold great potential for habitat/species protection and advancing interdisciplinary scientific research. Recently, numerous large-scale bird audio datasets created through crowdsourcing have played a key role in increasing research efforts~\cite{lehikoinen2023successful,xeno2023dataset}. These datasets facilitate the development and deployment of data-driven models aimed at solving the challenge of bird audio classification~\cite{stowell2017bird}. 
Whilst many bespoke models have been proposed for bird sound classification, we note that training these from scratch is particularly label inefficient~\cite{hockman2018acoustic}.
Therefore, we see that pretrained models are increasingly used to accelerate research by acting as well-trained feature extractors. These general models are tailored to different domains through fine-tuning and parameter-efficient classification heads~\cite{wang2023large}. For example, pretrained models from computer vision (e.g. ResNet~\cite{he2016deep}) show reasonable out-of-the-box classification performance by treating simple spectrographic features as images. Conversely, in the general acoustic analysis field, some recent works like LEAF~\cite{zeghidour2021leaf} and BEATs~\cite{chen2022beats} have made substantial progress in learning features directly from raw waveforms without any pre-transformation.

\begin{figure}[!t]
\centering
\includegraphics[width=1.0\columnwidth]{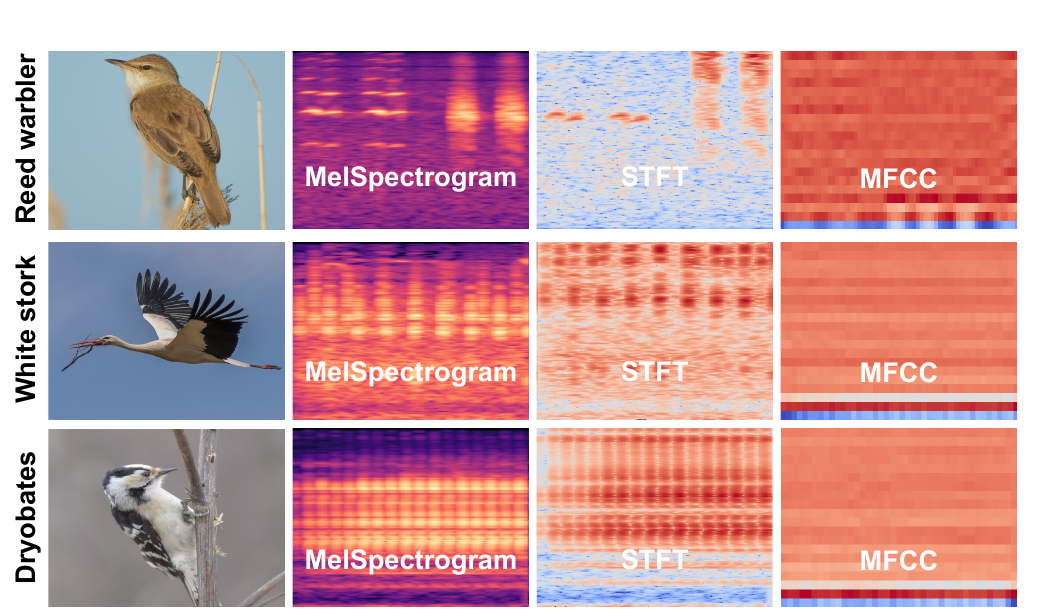}
\vspace{-1.5em}
\caption{Visualization examples of bird sound spectral features, including MelSpectrogram, STFT and MFCC.} 
\vspace{-2.1em}
\label{fig3_physical}
\end{figure}

Although the above datasets and algorithms have contributed to bird sound classification tasks, it is not immediately clear which approach to follow to build an efficient (both in terms of training sample size and model parameter count) and accurate model, with low effort. Bird sound classification is a specialized sub-field of acoustic classification which has significant challenges - a species may have a wide repertoire of vocalizations, yet may sound quite similar to another species. This requires highly discriminative feature extraction, yet we require our approach to be label-efficient, due to the high cost of annotating extensive datasets~\cite{wang2022efficient}. This contradiction can be resolved to an extent by using pretrained models, but we first need to transform the raw acoustic data to a more compact representation. To perform this transformation, we note that there is an inherent tension between spectral feature extraction and learnable feature extraction, with each having different relative merits. On the one hand, using spectral features such as a MelSpectrogram (MEL), Short-Time Fourier Transform (STFT) or Mel-Frequency Cepstral Coefficient (MFCC) is well grounded in acoustic signal processing. Yet, even choosing which one of these transformations and its parameters involves significant effort and tuning. On the other hand, learnable features such as BEATs or LEAF operate directly on the raw waveform and are well-trained on large corpora of general acoustic data. However, its direct use for bird sound classification remains to be investigated. Overall, while spectral feature-based methods currently achieve better classification performance, pretrained models are more sample-efficient and parameter-efficient and can yield diverse representation outcomes.

\begin{figure*}[!t]
\centering
\includegraphics[width=1\textwidth]{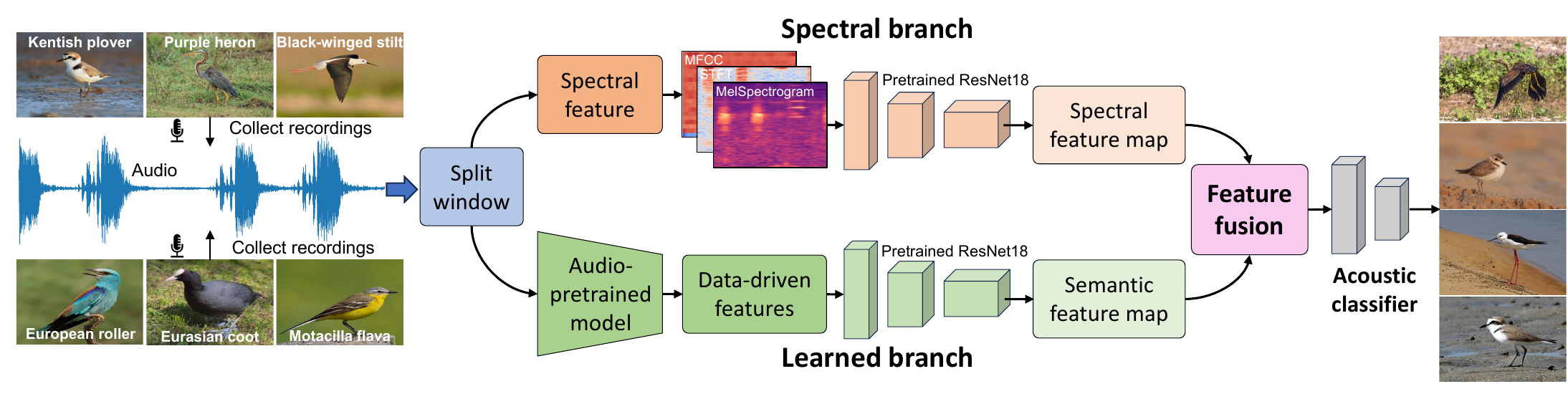}
\vspace{-2em}
\caption{The architecture of the proposed SSL-Net for efficient bird sound classification. It is made up of a \textcolor{LimeGreen}{\textbf{learned branch}} constructed using the audio-pretrained model and a \textcolor{YellowOrange}{\textbf{spectral branch}} built on various acoustic spectral representations. For the \textcolor{Lavender}{\textbf{feature fusion}}, we present three feasible and selectable strategies, the specifics are illustrated in Fig. \ref{fig2_fusion}. Moreover, a lightweight \textcolor{gray}{\textbf{acoustic classifier}} is used for the final sound classification.} 
\vspace{-1em}
\label{fig1_workflow}
\end{figure*}

In this study, we propose an efficient \textbf{\underline{S}}ynergistic \textbf{\underline{S}}pectral and \textbf{\underline{L}}earning-based \textbf{\underline{Net}}work called \textbf{SSL-Net} to handle the challenges bird sound classification. Instead of attempting to extensively tune models and architectures for maximum performance, we show how to synergize multiple approaches to yield efficient and accurate classification.
By using spectral representations to compensate for the distribution bias of learned features, our method achieves leading results. Importantly, it does so with low practitioner effort, low model complexity and high sample efficiency. Specifically, our main contributions are:

\vspace{-0.5em}
\begin{itemize}
    \item We present an efficient bird sound classifier based on synergistic spectral and learned information, which uses spectral representations to mitigate distribution bias in the pretrained model features and enhances performance and robustness with minimal labels.
    \vspace{-0.5em}
    \item We put forth different fusion strategies to mitigate the distribution bias within spectral and learned features, thereby yielding enhanced feature representations for the classification task.
    \vspace{-0.5em}
    \item Through comprehensive evaluations and ablation experiments with real-world bird audio data, our framework aids engineers in designing efficient and accurate classifiers with low effort.
\end{itemize}
\vspace{-0.8em}

\vspace{-0.85em}
\section{Related work}
\label{sec:Related_work}
\vspace{-0.3em}
AI-based bird sound classifiers have been extensively researched. Many of them leverage acoustic characteristics (e.g., MEL, STFT, and MFCC shown in Fig.~\ref{fig3_physical}) as feature maps, subsequently integrating them as images and using machine learning-based models for classification. For example, Xie and Zhu~\cite{xie2019handcrafted} proposed to use handcrafted features combined with deep learning to classify 14 bird sounds and achieved good performance. Besides, Tuncer \textit{et al.}~\cite{tuncer2021multileveled} presented a multileveled and handcrafted features-based machine learning model and good results in a dataset of 18 categories. Moreover, Nanni \textit{et al.}~\cite{nanni2021ensemble} and Jasim \textit{et al.}~\cite{jasim2022classify} designed different enhanced convolutional neural networks with spectral features to improve the model classification ability as well as robustness respectively. There are also some related and similar researches~\cite{saad2022classification, noumida2022multi, rajan2022bird, xie2022sliding}.

In addition, the audio-pretrained models developed over the past two years (e.g., Panns~\cite{kong2020panns}, LEAF~\cite{zeghidour2021leaf} and BEATs~\cite{chen2022beats}) are different from the above-mentioned methods. They rely on data-driven mining of representations that deviate from the traditional spectral acoustic features and have a wide range of applications, including acoustic feature extraction, audio classification, emotion analysis, keyword spotting, etc. Therefore, it is possible to efficiently fine-tune the generic audio-pretrained model in the bird sound classification tasks.

\vspace{-0.85em}
\section{Methodology}
\label{sec:Methodology}
\vspace{-0.2em}
In this section, we introduce the proposed SSL-Net as illustrated in Fig.~\ref{fig1_workflow}. It comprises two branches: the learned branch and the spectral branch, integrated through a feature fusion module. The former leverages an audio-pretrained model followed by an encoder that generates a semantic feature map. The latter utilizes acoustic features in conjunction with an encoder to derive a spectral feature map. Within the feature fusion module, we explore three distinct fusion strategies and apply them to merge the feature maps obtained from both branches for the final classification task.

\vspace{-0.4em}
\subsection{Learned branch} \label{subsec:sourcebranch}
The learned branch focuses on leveraging audio-pretrained models for general acoustic feature extraction. It is composed of two components: an audio-pretrained model and a pretrained model-based encoder. This approach serves two purposes: 1) enhancing the efficiency of feature extraction while 2) concurrently diminishing the required training sample size. We use BEATs~\cite{chen2022beats} as the backbone of the acoustic pretrained model. It extracts data-driven representations $\mathbf{a_{sem}}$ from split original audio data $\mathbf{x}$, as $\mathbf{a_{sem}} = \mathsf{BEATs}(\mathbf{x})$. After resizing representation $\mathbf{a_{sem}}$ to the same size, the pretrained ResNet18~\cite{he2016deep} backbone is used as the encoder for further feature mining and to obtain the semantic feature map $\mathbf{f_{sem}}$.

Since the full module of this branch uses pretrained models, it is very efficient without adding more weights. However, the semantic feature map is biased for bird sounds because it is pretrained using other general species' sounds dataset.

\vspace{-0.4em}
\subsection{Spectral branch} \label{subsec:targetbranch}
As the learned branch primarily encapsulates information from the general sound domain, it inherently lacks insights into the target domain (i.e., bird sounds). To mitigate it, we introduce the spectral branch which serves to compensate for the overall distribution by incorporating information pertinent to the target domain. It includes two modules: a spectral features extractor and a pretrained encoder. The former extractor processes the segmented audio signal, extracting key acoustic attributes based on parameters suitable for bird audio, namely the MEL, STFT, and MFCC. Subsequently, these three features are resized to uniform dimensions and concatenated as individual channels. This composite representation $\mathbf{a_{spe}}$ is then input into a pretrained ResNet18~\cite{he2016deep}  for further extraction. The output of the last convolutional layer is used as the spectral feature map $\mathbf{f_{spe}}$.

Similar to the first branch, the spectral branch exclusively employs pretrained models and incorporates acoustic spectral features that require no iterative training. Consequently, this entire branch operates with remarkable efficiency. Furthermore, it can mitigate the biased distribution in the preceding branch within the target domain. More details about the branches will be shown in the ablation experiment in Sec.~\ref{sec:Experiment}.

\subsection{Feature fusion module} \label{subsec:fusion}
\begin{figure}[!t]
\centering
\includegraphics[width=1.0\columnwidth]{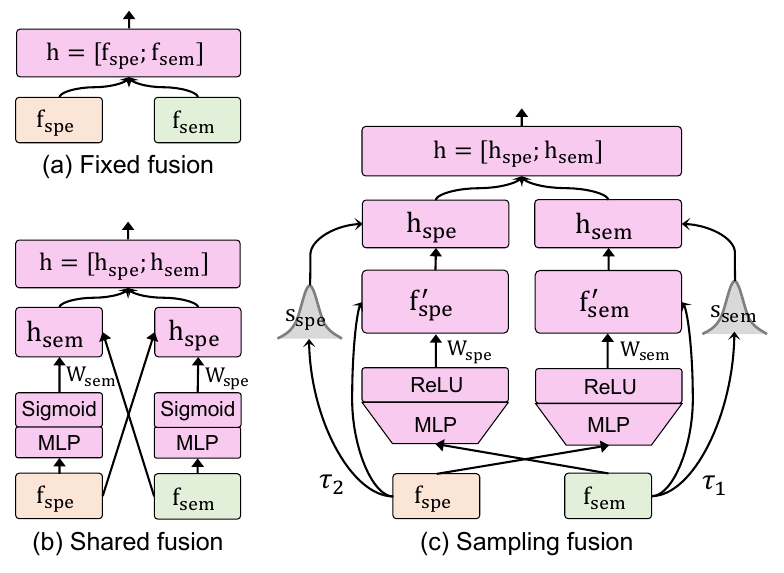}
\vspace{-2em}
\caption{Flow charts of the proposed fixed, shared, and sampling feature fusion strategies.} 
\label{fig2_fusion}
\vspace{-1em}
\end{figure}

Building on top of the extracted semantic feature $\mathbf{f_{sem}}$ and spectral feature $\mathbf{f_{spe}}$ from the two branches, we design three feature fusion strategies with the same input and output, as shown in Fig.~\ref{fig2_fusion}, to facilitate the transfer from the general acoustic domain to the target domain and merge the complementary features effectively.

\textbf{1) Fixed fusion} is a straightforward approach, as shown in Fig.~\ref{fig2_fusion}(a). Two representations $\mathbf{f_{sem}}$ and $\mathbf{f_{spe}}$ are directly concatenated into a new fusion feature $\mathbf{h}$. This most efficient fusion can be modeled as $\mathbf{h} = \mathsf{Concat}[\mathbf{f_{spe}}; \mathbf{f_{sem}}]$.

\textbf{2) Shared fusion} is similar to the attention mechanism~\cite{vaswani2017attention}, as shown in Fig.~\ref{fig2_fusion}(b). This strategy enables the fusion process to be simultaneously trained alongside other modules by conditioning and reweighting both features concurrently. First, we perform the following operations on the semantic feature $\mathbf{f_{sem}}$ and spectral feature $\mathbf{f_{spe}}$.
\begin{equation}
\small
    \mathbf{W_{sem}} = \mathsf{Sigmoid}(\mathsf{Linear}(\mathbf{f_{spe}})),
\end{equation}
\begin{equation}
\small
    \mathbf{W_{spe}} = \mathsf{Sigmoid}(\mathsf{Linear}(\mathbf{f_{sem}})),
\end{equation}
\noindent where $\mathsf{Linear}(\cdot)$ denotes a multilayer perceptron (MLP). $\mathbf{W_{sem}}$ and $\mathbf{W_{spe}}$ are the new fusion weights. Sigmoid is used to ensure the weights are in the range of 0 to 1. The next step is to perform element-wise multiplication of the weights with the features and get representations $\mathbf{h_{sem}}$ and $\mathbf{h_{spe}}$. The two representations are concatenated into the final feature $\mathbf{h}$.
\begin{equation}
\small
    \mathbf{h} = \mathsf{Concat}[\mathbf{f_{spe}} \odot \mathbf{W_{spe}};\mathbf{f_{sem}} \odot \mathbf{W_{sem}}].
\end{equation}

\textbf{3) Sampling fusion} refers to learning a stochastic function that samples in two features, as shown in Fig.~\ref{fig2_fusion}(c). Compared with the shared fusion, it could achieve more robust feature fusion. Firstly, we implement the operations on the semantic feature $\mathbf{f_{sem}}$ and spectral feature $\mathbf{f_{spe}}$.
\begin{equation}
\small
    \mathbf{f'_{sem}} = \mathbf{f_{sem}} \odot \mathsf{ReLU}(\mathsf{Linear}(\mathbf{f_{spe}})),
\end{equation}
\begin{equation}
\small
    \mathbf{f'_{spe}} = \mathbf{f_{spe}} \odot \mathsf{ReLU}(\mathsf{Linear}(\mathbf{f_{sem}})),
\end{equation}

\noindent where the $\mathsf{Linear}(\cdot)$ will boost the dimensions to double for the following distribution operations. 

Next, the original features, $\mathbf{f_{sem}}$ and $\mathbf{f_{spe}}$, are individually sampled using the Gumbel-Softmax (GS) distribution~\cite{jang2016categorical}. It is selected due to its differentiability during the sampling procedure. This property enables backpropagation, thus facilitating end-to-end model training. 
\begin{equation}
\small
    \mathbf{s_{sem}} \sim p(\mathbf{s_{sem}}|\mathbf{f_{sem}}, \tau_1) = \mathsf{GS}(\mathbf{f_{sem}}),
\end{equation}
\begin{equation}
\small
    \mathbf{s_{spe}} \sim p(\mathbf{s_{spe}}|\mathbf{f_{spe}}, \tau_2) = \mathsf{GS}(\mathbf{f_{spe}}),
\end{equation}
where $\tau_1$ and $\tau_2$ represent the temperature parameter in GS distribution. Finally, we multiply the GS distribution with new representations correspondingly and get the final fusion feature representations $\mathbf{h_{sem}}$ and $\mathbf{h_{spe}}$, which is: 
\begin{equation}
\small
    \mathbf{h} = \mathsf{Concat}[\mathbf{f'_{spe}} \odot \mathbf{s_{spe}}; \mathbf{f'_{sem}} \odot \mathbf{s_{sem}}].
\end{equation}

\subsection{Final acoustic classifier} \label{subsec:classifier}
We apply a lightweight MLP to classify the fused features and get the final classification score. The iterative training process is performed using the cross-entropy loss function.

\section{Experiments}
\label{sec:Experiment}

In this section, we validate the proposed SSL-Net using the Western Mediterranean Wetland Birds (WMWB) dataset~\cite{gomez2022western} collected from Xeno-canto~\cite{xeno2023dataset}, which consists of 20 classes over 200 minutes. We also perform full ablation experiments and trade-offs to evaluate each module in the pipeline.

\begin{table}[!t]
   \centering
   \caption{The effectiveness of different single branch methods with WMWB dataset~\cite{gomez2022western}. The results at * are different due to the different classifier with the same number of parameters.}
   \vspace{0.2em}
   \renewcommand{\arraystretch}{1}
   \resizebox{1\columnwidth}{!}{
\begin{tabular}{ccc|cccc|c}
\hline \hline
\multirow{2}{*}{\textbf{Branch}} & \multirow{2}{*}{\textbf{Model}} & \multirow{2}{*}{\textbf{Feature}} & \multicolumn{4}{c|}{\textbf{Metric}} & \multirow{2}{*}{\textbf{Param (\#)}} \\
 &  &  & Acc$\uparrow$ (\%) & Prc$\uparrow$ (\%) & Rec$\uparrow$ (\%) & F1$\uparrow$ &  \\ \hline
\multirow{2}{*}{Baseline~\cite{gomez2022western}} & ResNet50 & MEL & 77.70 & 74.68 & 72.33 & 73.60 & 542,292 \\ \cline{8-8} 
 & ResNet18 & MEL$^*$ & 76.10 & 72.23 & 71.98 & 72.42 & \multirow{7}{*}{68,500} \\ \cline{1-7}
\multirow{2}{*}{Learned} & \multirow{2}{*}{ResNet18} & LEAF & 83.60 & \textcolor{red}{\textbf{86.16}} & 77.14 & 78.50 &  \\
 &  & BEATs & 83.23 & 80.56 & 78.61 & 79.02 &  \\ \cline{1-7}
\multirow{4}{*}{Spectral} & \multirow{4}{*}{ResNet18} & MEL$^*$ & 83.68 & 81.52 & 74.43 & 76.78 &  \\
 &  & STFT & 83.45 & 83.25 & 73.98 & 74.99 &  \\
 &  & MFCC & 83.31 & 81.17 & 72.36 & 74.26 &  \\
 &  & ALL & \textcolor{red}{\textbf{84.02}} & 83.11 & \textcolor{red}{\textbf{82.70}} & \textcolor{red}{\textbf{81.63}} &  \\ \hline \hline
\end{tabular}}
\vspace{-1em}
    \label{table:1}
\end{table}
Tabel~\ref{table:1} presents the results from the single branch methods, where 200 labeled samples per species are used. The learned branch methods and spectral branch methods outperform the baseline (ResNet50) with fewer learnable parameters. Besides, the learned branch achieves better performance than a single spectral feature, which demonstrates the adaptability and transferability of the generic audio-pretrained model to the fine-grained task of bird sound classification. In addition, the spectral branch utilizing three features outperforms the methods solely from the learned branch, as spectral features are better suited for the target domain (bird sound).

\begin{table}[!t]
   \centering
   \caption{The effectiveness of different fusion strategies with different branches. We use accuracy, precision, recall and F1-score as evaluation metrics.}
   \vspace{0.2em}
   \renewcommand{\arraystretch}{1}
   \resizebox{1.0\columnwidth}{!}{
\begin{tabular}{cc|cccc|c}
\hline \hline
\multirow{2}{*}{\textbf{Fusion}} & \multirow{2}{*}{\textbf{Branch}} & \multicolumn{4}{c|}{\textbf{Metric}} & \multirow{2}{*}{\textbf{Param (\#)}} \\
 &  & Acc$\uparrow$ (\%) & Prc$\uparrow$ (\%) & Rec$\uparrow$ (\%) & F1$\uparrow$ &  \\ \hline
\multirow{3}{*}{Fixed} & LEAF+BEATs & 83.48 & 85.34 & 79.06 & 79.88 & \multirow{3}{*}{134,036} \\
 & Spectral+LEAF & 83.33 & 87.78 & 84.24 & 83.71 &  \\
 & Spectral+BEATs & 83.65 & 86.92 & 86.28 & 86.12 &  \\ \hline
\multirow{3}{*}{Shared} & LEAF+BEATs & 84.62 & 85.59 & 80.89 & 81.90 & \multirow{3}{*}{2,260,964} \\ 
 & Spectral+LEAF & \textcolor{red}{\textbf{87.29}} & 90.38 & 85.81 & 86.53 &  \\
 & Spectral+BEATs & 86.54 & 89.60 & 88.29 & 88.68 &  \\ \hline
\multirow{3}{*}{Sampling} & LEAF+BEATs & 83.37 & 84.79 & 80.03 & 81.59 & \multirow{3}{*}{4,262,708} \\ 
 & Spectral+LEAF & 86.06 & \textcolor{red}{\textbf{91.98}} & 86.14 & 87.48 &  \\
 & Spectral+BEATs & 85.70 & 89.69 & \textcolor{red}{\textbf{88.77}} & \textcolor{red}{\textbf{88.79}} &  \\ \hline \hline
\end{tabular}}
   \vspace{-1.5em}
    \label{table:2}
\end{table}
Table~\ref{table:2} shows the effectiveness of different fusion strategies with
different branches. The majority of BEATs+Spectral-based methods achieved optimal results, which shows the effectiveness of the BEATs as a generic audio-pretrained model. However, the results of BEATs+LEAF (fusing two learned branches) are comparatively subpar, due to both representations being data-driven general acoustic representations, which introduces distributional bias in the target domain. As for fusion strategies, the sampling fusion excels in classification performance, but the shared fusion, while slightly sacrificing classification performance, reduces trainable parameters and model complexity, occupying the efficiency-classification performance trade-off peak. Lastly, the simplest but quite efficient fixed fusion, using the fewest parameters, also achieved good results. In summary, three fusion strategies effectively synergize distinct representations from different domains, resulting in state-of-the-art outcomes.

\begin{figure}[!t]
\centering
\includegraphics[width=0.89\columnwidth]{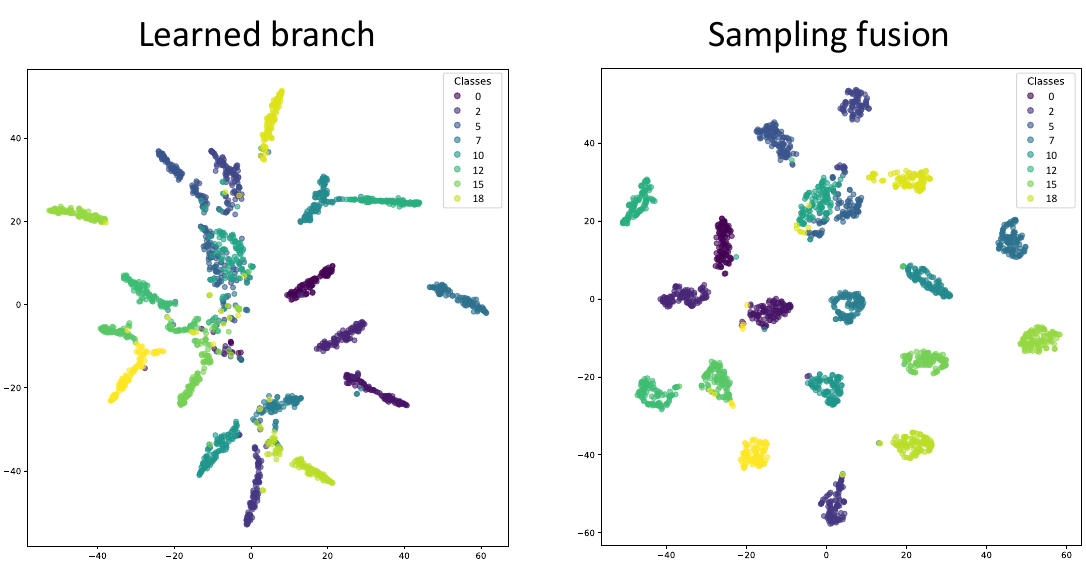}
\vspace{-1em}
\caption{Visualisation of the single branch features and fused features before final classification using t-SNE.} 
\vspace{-0.2em}
\label{fig4_tsne}
\end{figure}
To further study the advantages of fusion, we visualize the pre-fusion (learned branch for example) and post-fusion features (sampling fusion for example), as shown in Fig~\ref{fig4_tsne}. Clearly, the fusion module yields more discriminative features, allowing for more efficient classification with fewer learnable parameters while ensuring robust classification.

\begin{figure}[!t]
\centering
\includegraphics[width=0.82\columnwidth]{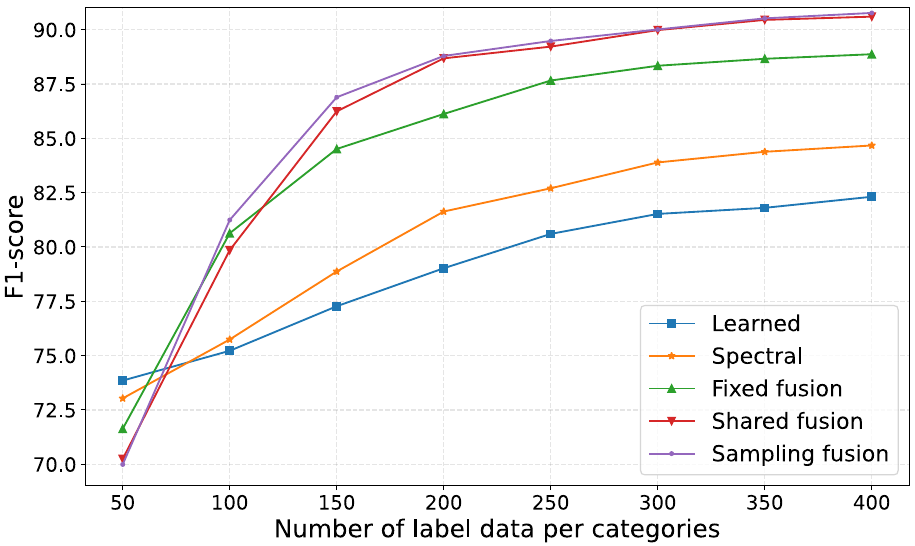}
\vspace{-1em}
\caption{The impact of the number of labeled samples on performance with different methods.} 
\vspace{-0.5em}
\label{fig5_label}
\end{figure}
Finally, we explore the impact of the sample quantity, as depicted in Fig~\ref{fig5_label}. The performance of all methods improved as the number of labeled samples increased. Among them, the fusion-based approaches show the most pronounced variation, ultimately achieving optimal performance with a larger quantity of samples. This suggests that our proposed methods can help researchers and engineers select appropriate models in the case of limited samples with low effort.

\vspace{-0.5em}
\section{Conclusion}
\label{sec:Conclusion}

In the paper, we propose a synergistic spectral and learning-based network (SSL-Net) to handle bird sound classification. Our approach with three diverse fusion strategies efficiently bridges the gap between generic audio and bird audio by leveraging spectral representations to address distribution bias in the pretrained model features. The results from the audio dataset of 20 bird species, collected in the real field, affirm both the effectiveness and efficiency of our method within the limitations of the labeled sample size.

\noindent \textbf{Note:} This work has been submitted to the IEEE for possible publication. Copyright may be transferred without notice, after which this version may no longer be accessible.
\bibliographystyle{IEEEbib}
\bibliography{6_reference}

\end{document}